\documentclass[a4paper]{jpconf}
\usepackage{graphicx}
\begin{document}
\title{Modeling the structure of magnetic fields in Neutron Stars: from the  interior to the magnetosphere}

\author{Niccol\`o Bucciantini$^{1,2,3}$, Antonio G. Pili$^{2,1,3}$ and Luca Del Zanna$^{2,1,3}$}

\address{$^1$ INAF - Osservatorio Astrofisico di Arcetri, L.go Fermi 5,
  50125, Firenze, Italy}
\address{$^2$ Univ. Firenze - Dip. di Fisica e Astronomia ,
  via G.Sansone 1,
 50019, Sesto F.no, Italy}
\address{$^3$ INFN - Sezione di Firenze, via G.Sansone 1, 50019, Sesto F.no, Italy}

\ead{niccolo@arcetri.astro.it}

\begin{abstract}
The phenomenology of the emission of pulsars and magnetars depends
dramatically on the structure and properties of their magnetic
field. In particular it is believed that the outbursting and flaring
activity observed in AXPs and SRGs is strongly related to their
internal magnetic field.  Recent observations have moreover shown that
charges are present in their magnetospheres supporting the idea that
their magnetic field is tightly twisted in the vicinity of the star.
In principle these objects offer a unique opportunity to investigate
physics in a regime beyond what can be obtained in the laboratory. We
will discuss the properties of equilibrium models of magnetized
neutron stars, and we will show how internal and external currents can
be related. These magnetic field configurations will be discussed
considering also their stability, relevant for their origin and possibly connected to
events like SNe and GRBs. We will also show what kind of 
deformations they induce in the star, that could lead to emission of
gravitational waves. In the case of a twisted magnetosphere we will
show how the amount of twist regulates their
general topology. A general formalism based on the simultaneous numerical solution of the general relativistic Grad-Shafranov equation and Einstein equations will be presented. 
\end{abstract}

\section{Introduction}

Neutrons Stars (NSs) are known to show a diverse phenomenology, that
allows us to separate them into several classes.  Among these classes, Anomalous
X-Ray Pulsars (AXPs) and Soft Gamma-ray Repeaters (SGRs) have
attracted attention because of their extraordinary energetic 
properties: a persistent X-ray emission 
with luminosities $L_X\sim  10^{33} - 10^{36}~\mbox{erg s}^{-1}$;
flaring activity with X-ray bursts lasting  $\sim 0.1 -1$~s and with peak luminosities 
$\sim 10^{40}-10^{41} \mbox{ erg s}^{-1}$, and in a few cases violent events, known as giant flares, 
during which an amount of energy $\sim 10^{44} - 10^{46}~\mbox{ erg}$ 
is released \cite{Mereghetti08a,Rea_Esposito11a,Turolla_Esposito13a}.

Today SGRs and AXPs are grouped in the  same class of NSs
called \textit{magnetars} \cite{Duncan_Thompson92a,Thompson_Duncan93a}. 
These are young (with a typical age of $10^4$~yr), 
isolated NSs with rotational period in the 
range  $\sim 2 -12$~s, and with a typical magnetic field in the range $10^{14} - 10^{15}$~G \cite{Kouveliotou_Strohmayer+99a}.  
Since they are slow rotators, spin-down energy losses  cannot power their emission, 
which is instead believed to come from the dissipation and
rearrangement of their magnetic energy. 

The magnetic field has important implication on the way NSs manifest themselves
in the electromagnetic spectrum. In particular for Pulsars, 
models of the outer magnetosphere have been developed since the '60
\cite{Goldreich_Julian68}, up to the present day
\cite{Tchekhovskoy_Spitkovsky13a}. 
On the other hand, the interest on the interior structure has been mostly driven
by questions of nuclear and theoretical physics, especially their
Equation of State (EoS) \cite{Chamel_Haensel08a,Lattimer12a} and their cooling
properties \cite{Yakovlev_Pethick04a,Yakovlev_Gnedin+05a}. 

The simultaneous presence of high density, strong gravity, and strong magnetic
fields makes magnetars a unique environment.
The study of the properties and geometry of the internal magnetic
field  in NSs is thus an important step towards a complete understanding of magnetars.
 The analysis of equilibrium
configurations has mainly focused on understanding the effects of 
the magnetic field on the structure of the star. Strong magnetic 
fields deform the star, and such deformations, in 
conjunction with fast rotation, could lead to emission of 
Gravitational Waves (GW) \cite{Mastrano_Larsky+13a,DallOsso_Stella07a}. 
Particular efforts have been recently aimed at investigating the
stability of various magnetic configurations. However due to the
complexity of the problem,
models have been worked either in Newtonian regime 
\cite{Lander_Jones09a,Glampedakis_Andersson+12a,Fujisawa_Yoshida+12a}, 
or in GR with a perturbative approach
\cite{Ciolfi_Ferrari+09a,Ciolfi_Ferrari+10a,Ciolfi_Rezzolla13a}, and
with currents purely confined to the interior.
Only recently \cite{Pili_Bucciantini+14a,Bucciantini_Pili+15a} they have been worked out in the fully
non-linear GR regime, including the possibility of currents extending
outside into a magnetosphere \cite{Pili_Bucciantini+15a}.

We present and review here a study conducted by our group, aiming at
modeling magnetized NSs in general relativity. Our approach is based
on the simultaneous solution of the Einstein equations for the GR
metric,  of the  general-relativistic Euler equation for the hydromagnetic equilibrium, and
the general-relativistic 
Grad-Shafranov equation for the magnetic field structure, using a
formalism that allows 
electric currents to flow outside the star. We have investigated
several new functional forms for the current distribution,
and studied the properties of the resulting magnetic field. 

\section{Solving Einstein equations}

Given a generic spacetime, 
the  line element can be written as \cite{Alcubierre08a,Gourgoulhon12a}:
\begin{equation}
ds^2 = -\alpha^2dt^2+ \gamma_{ij}(dx^i+\beta^i dt)(dx^j+\beta^j dt),
\end{equation}
where $\alpha$ is called the \emph{lapse} function, $\beta^i$ is the
\emph{shift vector}, $\gamma_{ij}$ is the \emph{three-metric}, and
$i,j=r,\theta,\phi$, if a
spherical coordinate system $x^\mu = (t, r, \theta, \phi) $ is chosen.  The assumptions
of \emph{stationarity} and \emph{axisymmetry} imply that all metric
terms are only a function of $r$ and $\theta$.

For neutron stars endowed with an axisymmetric magnetic field the
metric can be approximated to a high degree of accuracy as {\it
  conformally flat}  \cite{Wilson_Mathews03}, such that in
spherical coordinates:
\begin{equation}
ds^2=-\alpha^2 dt^2+ \psi^4 (dr^2 + r^2d\theta^2 + r^2\sin{\theta}^2d\phi^2),
\end{equation}
where $\psi$ is the conformal
factor, and for simplicity we have neglected  frame dragging given that magnetar
are typically slow rotators.

The conformally flat condition (CFC) allows us to cast Einstein's
equations in a simpler and numerically stable form and, as a 
consequence, it can handle stronger fields and deformations without compromising the 
accuracy of the results as we will discuss.
With this approximation Einstein's equations reduce 
to the following Poisson-like equations:
\begin{equation}
\Delta \psi =-[2\pi\psi^6(e +B^2/2)]\psi^{-1},
\label{eq:1}
\end{equation}
\begin{equation}
\Delta(\alpha\psi)=[2\pi\psi^6(e +B^2/2) +2\pi\psi^6(6p+B^2)\psi^{-2}](\alpha\psi),
\label{eq:2}
\end{equation}
where $\Delta$ is the standard Laplacian operator in spherical coordinate while $e$, 
$p$ and $B^2$ are respectively the energy density, the pressure and 
the magnetic field energy density as measured in the lab frame. 

For the non-linear Poisson-like equations
Eq.~\ref{eq:1}-\ref{eq:2} we employ the XNS
algorithm
\cite{Bucciantini-Del_Zanna11a,Del-Zanna_Zanotti+07a,Bucciantini_Del-Zanna13a}. The
code is designed to compute hydromagnetic equilibria even in the
presence of rotation.
 Axisymmetric solutions are searched
in terms of a series of spherical harmonics $Y_l(\theta)$:
\begin{equation}
q(r,\theta):=\sum_{l=0}^{\infty}[A_l(r)Y_l(\theta)].
\end{equation}
The Laplacian can then be reduced to a series of radial 2nd order
boundary value ODEs for the
coefficients $A_l(r)$ of each harmonic, which are then solved using
tridiagonal matrix inversion.  This procedure is repeated until convergence, using in the
source term the value of the solution computed at the previous
iteration. We found that 20 spherical harmonics for the
elliptic solvers and a grid in spherical coordinates in the
domain $r=[0,30]$km, $\theta=[0,\pi]$, with 250 points in the radial
direction and 100 points in the angular one, are sufficient to achieve
a numerical accuracy  of the order of $10^{-3}$. Comparison with results already
present in literature for fast unmagnetized rotators, shows that the
CFC approximation is correct to an accuracy $\sim 10^{-3}$ even for
highly deformed stars.

\section{Magnetic equilibria}

The only  non-vanishing equation of the static GRMHD system 
is the Euler equation in the presence of an external electromagnetic field:
\begin{equation}
\partial_i p + (e \! + \! p) \,\partial_i \ln\alpha = L_i := \epsilon_{ijk} J^j B^k,
\label{eq:lorentz}
\end{equation}
with $i=r,\theta$, and where $L_i$ is the Lorentz force and $J^i = \alpha^{-1}\epsilon^{ijk}\partial_j (\alpha B_k)$
is the conduction current. Assuming a \emph{barotropic} EoS
$p=p(\rho)$, $e=e(\rho)$ ($\rho$ is the rest mass density), we find
\begin{equation}
\partial_i \ln h + \partial_i \ln\alpha  = \frac{L_i}{\rho h},
\end{equation}
where the specific enthalpy  is $h:= (e+p)/\rho$.
Integrability requires the
existence of a scalar function $\mathcal{M}$
such that $L_i=\rho h \partial_i\mathcal{M}$. 

In the case of a purely toroidal field, the Lorentz force is conveniently written in terms of 
$\alpha B_\phi$, and the Euler equation reads
\begin{equation}
\partial_i \ln h + \partial_i \ln \alpha + 
\frac{\alpha B_\phi\partial_i ( \alpha B_\phi ) }{ \rho h \, \alpha^2 \psi^4r^2\sin^2\!\theta} = 0,
\end{equation}
such that
\begin{equation}
B_\phi = \alpha^{-1} \mathcal{I}(\mathcal{G}), \quad
\mathcal{M}(\mathcal{G}) = - \int \frac{\mathcal{I}}{\mathcal{G}}\frac{d\mathcal{I}}{d
  \mathcal{G}}d\mathcal{G},\quad {\rm with} \quad \mathcal{G} = \rho h \, \alpha^2 \psi^4r^2\sin^2\!\theta.
\end{equation}
Among all possible functional choices we have selected a {\it magnetic
  barotropic law}:
\begin{equation}
\mathcal{I}(\mathcal{G}) = K_m \mathcal{G}^m, \quad \mathcal{M}(\mathcal{G}) = - \frac{m
  K^2_m}{2m-1} \mathcal{G}^{2m-1}.
\label{eq:bernoullitor}
\end{equation}

In the case where a poloidal magnetic field
is present  a formulation based on the so-called
\emph{Grad-Shafranov equation}
for the toroidal component of the vector potential $A_\phi$ is more
convenient. In this case:
\begin{equation}
L_i = \rho h\, \partial_i \mathcal{M} = \rho h\, \frac{d\mathcal{M}}{d
  A_\phi}\partial_i A_\phi \rightarrow
\ln{\left( \frac{h}{h_c}\right)} + \ln{\left( \frac{\alpha}{\alpha_c}\right)} - \mathcal{M}(A_\phi) = 0,
\label{eq:bernoulli}
\end{equation}
where constants are calculated at the stellar center.
Moreover, the $\phi$ component of the Lorentz force,
which must also vanish, implies
\begin{equation}
B_\phi = \alpha^{-1} \mathcal{I}(A_\phi).
\end{equation}
Introducing $\sigma := \alpha^2 \psi^4 r^2\sin^2\!\theta$ and $\tilde{A}_\phi := A_\phi / (r\sin\theta)$ 
and the new operator 
\begin{equation}
\tilde{\Delta}_3  \! := \!  \Delta - \frac{1}{r^2\sin^2\!\theta}  \! = \! 
\partial^2_r + \frac{2}{r}\partial_r+\frac{1}{r^2}\partial_\theta^2
+ \frac{1}{r^2\tan{\theta}}\partial_\theta - \frac{1}{r^2\sin^2\!\theta},
\end{equation}
for which $\tilde{\Delta}_3 \tilde{A}_\phi = \Delta_* A_\phi / (r\sin\theta)$
(it coincides with the $\phi$ component of the \emph{vector laplacian} in spherical coordinates),
after some calculations we retrieve the \emph{Grad-Shafranov} equation for the magnetic flux function $A_\phi$:
\begin{equation}
\tilde{\Delta}_3 \tilde{A}_\phi
+  \frac{\partial A_\phi \partial\ln (\alpha\psi^{-2})}{r \sin\theta}
+ \psi^8 r \sin\!\theta \left( \rho h \frac{d \mathcal{M}}{d  A_\phi}
+ \frac{\mathcal{I}}{\sigma}\frac{d\mathcal{I}}{dA_\phi} \right) = 0.
\label{eq:gs}
\end{equation}
Again we have investigated several functional forms for the scalar
functions $\mathcal{M}$ and $\mathcal{I}$, including non-linear terms
that allow us to model different current distributions inside the
star. In principle one can set the function $\mathcal{I}$ to vanish
outside the star, leading to the so called {\it twisted torus
  configurations}, where a twisted magnetic flux rope is present
inside the star. On the other hand if one allows $\mathcal{I}$ to be
non-zero also outside the star, one finds models with a so called
{\it twisted magnetosphere}.

 Interestingly the Grad-Shafranov equation
Eq.~\ref{eq:gs} can be reduced
to the solution of a non-linear vector Poisson equation, which is
formally equivalent to the elliptic equations that one needs to solve in
the CFC approximation. It is thus possible to use the same algorithm, with a combination of
vector spherical harmonics decomposition for the angular part, and
matrix inversion for the radial part, that was used in the metric
solver \cite{Bucciantini-Del_Zanna11a}.

\section{Results}

We begin by describing the results for purely toroidal magnetic
fields. By modifying the currents profile it is possible to obtain
different configurations where the magnetic field can be concentrated
either toward the center or toward the edge of the star. In
Fig.~\ref{fig:fig1} we show the outcome for different magnetic field
configurations. In all cases the maximum strength of the magnetic field is the
same as well as the mass of the Neutron Star. However it is
immediately evident that the deformation of the star is quite
different, being much stronger for fields concentrated at the
center. Given that a toroidally dominated configuration is expected to
result from the core-collapse of a rapidly rotating core, due to
differential rotation during infall, our results show that the
important parameter is not just the global strength of the differential
rotation, but its profile, which can be related to the pre-collapse
evolution of the progenitor.

\begin{figure}[h]
\begin{minipage}{39pc}
\includegraphics[width=39pc]{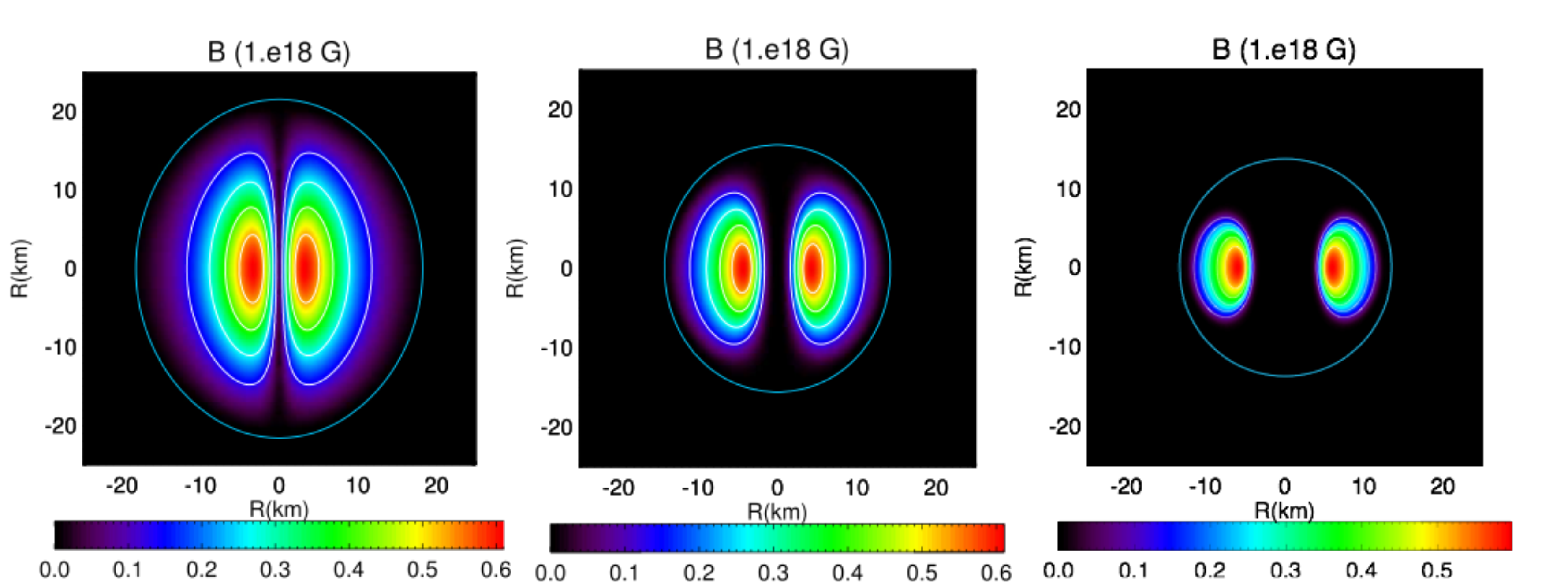}
\caption{\label{fig:fig1}Neutron stars with purely toroidal
  field. Paneles show the strength of the magnetic field for a star of
  mass $M=1.68M_\odot$. From left to right $m=1,2,4$. The blue line is
the stellar surface.}
\end{minipage} 
\end{figure}

In general for purely toroidal models we find that, as the magnetic
field increases the star inflates and its radius grows. Interestingly
this leads to a saturation of the maximum value that the magnetic
field can reach inside the star. Increasing the magnetization leads to a
bigger star and not a stronger field. We also find that the effect of
the magnetic field is more pronounced in low mass stars, while more
compact configurations, of higher central density, tend to show smaller
deformations, but can support strongher fields.

Models with purely poloidal field instead tend to show an oblate
deformation. Again it is possible to modify the current distribution
to achieve configurations where the field is distributed
differently. Interestingly now, for fields concentrated toward the edge
of the star, the deformation tends to be larger than in the case of
magnetic field concentrated toward the center. It is also possible to
obtain configurations where the magnetic field at the surface can
differ substantially from a simple dipole shape. By adopting diferent
prescriptions, we can make it either higher at
the equator, or concentrated toward the
axis, in a configuration where the bulk of the star is
demagnetized. Such configurations are reminiscent of recent full
multidimensional results in core-collapse simulations, where
demagnetized neutron stars are obtained due to turbulent flux
expulsion \cite{Obergaulinger_Janka+14a}. For very strong magnetic fields it is possible to obtain
even configurations where the density maximum is displaced from the
center due to magnetic field support.

\begin{figure}[h]
\begin{minipage}{39pc}
\includegraphics[width=39pc]{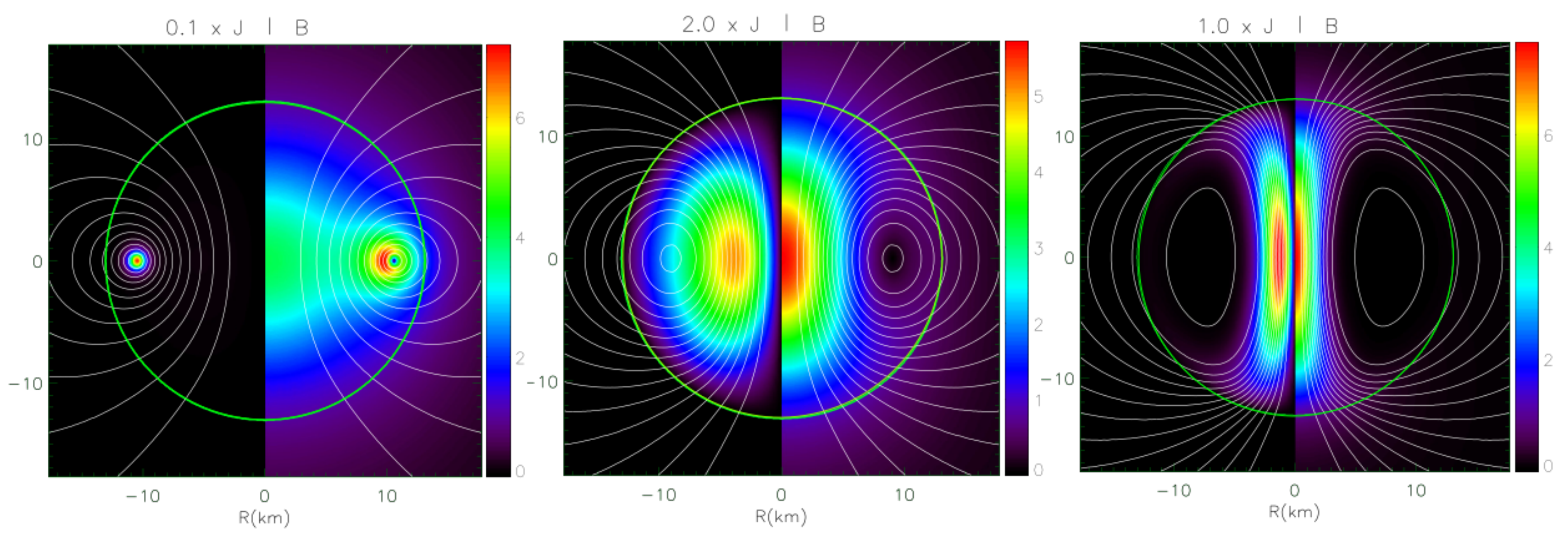}
\caption{\label{fig:fig1}Neutron stars with purely poloidal
  field. Panels show on the right the strength of the magnetic field
  (units $10^{14}$G),
  and on the left the strength of the currents, for a star of
  mass $M=1.68M_\odot$. The left panel shows the effect of additive
  non-linear currents that concentrate the field toward the surface,
  the right one the effect of subtractive currents that concentrate
  the field toward the axis. The blue line is
the stellar surface.}
\end{minipage} 
\end{figure}

We have computed Twisted-Torus configurations in the non-perturbative
regime. These  mixed configurations are favored based on stability
arguments.  The toroidal component can reach a strength comparable
with the poloidal one but it is always energetically subdominant. The
deformations are almost completely due to the poloidal field, acting on
the interior. Several functional forms for the current distribution
were used, but the system was always found to saturate to
configurations where the energy of the toroidal component is at most
10\% of the total magnetic energy. The main reason for this is that to
increase the toroidal magnetic field, one needs to rise also the
toroidal currents (the two being related due to equilibrium requirement
in the GS equation), which act to change the structure of the poloidal field,
reducing the volume occupied by the toroidal field. It is possible
to overcome this problem only by sacrificing global dynamical
equilibrium in the outer layers of the star, or in the case of
non-barotropic EoS. Interestingly it looks like such
energy ratio depends more on the stratification of the NS than on the
current distribution.

More recenlty \cite{Pili_Bucciantini+15a} models were built with currents extending also outside
the neutron star into a twisted magnetosphere. In the case of
magnetars there is an increasing set of observational evidence
pointing to the fact that their magnetosphere is 
endowed with a highly twisted magnetic field.
This is strongly suggested by the features of their persistent X-ray spectra, which
are well fitted by a blackbody-like component at $kT\sim 0.5$ keV,
likely thermal emission from the neutron star surface, joined 
with a  power-law tail that becomes dominant above 10 keV \cite{Kuiper_Hermsen+2006a}. 
The latter  can be explained in terms of resonant cyclotron
scattering of the thermal photons by magnetospheric particles 
\cite{Thompson_Lyutikov+02a,Beloborodov_Thompson07a}. Typically, in the standard reference model
\cite{Thompson_Lyutikov+02a} the magnetosphere 
is described in terms of a self-similar, globally twisted, dipolar magnetic field. 
This model has been refined to account for higher order multipoles 
\cite{Pavan_Turolla+09a}, in response to  observational indications of 
a local, rather than global, twist in the magnetosphere 
\cite{Woods_Kouveliotou+07a,Perna_Gotthelf08a}. 
Recently this scenario has been strengthened also by the detection
of a proton cyclotron feature in the X-ray spectrum of the
``low-field'' magnetar SGR 0418+5729 which is compatible with
a strong, but localized, toroidal field of the order
of $10^{15}$ G \cite{Tiengo_Esposito+13a}.

\begin{figure}[h]
\begin{minipage}{39pc}
\includegraphics[width=39pc]{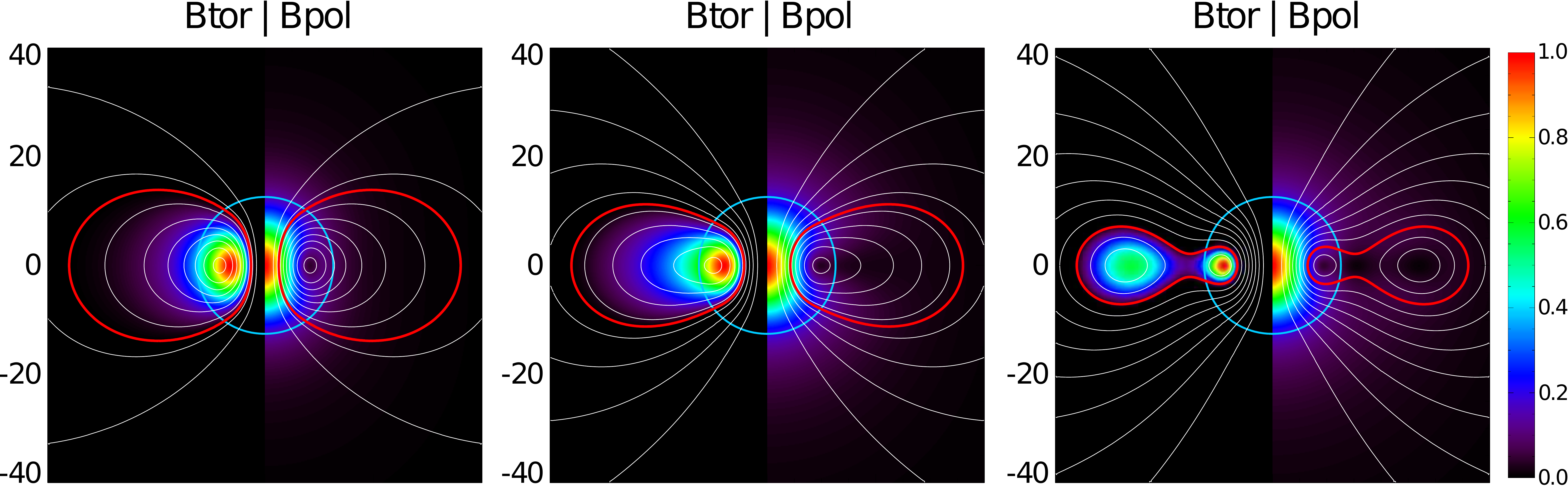}
\caption{\label{fig:fig1}Neutron stars with twisted
  magnetospheres. Panels show on the left the strength of the toroidal
component of the magnetic field, and on the right the strength of the
poloidal component (normalized to the maxima). From left to right models have progressively
higher level of twist, showing the transition from a configuration of
topologically connected magnetic regions to a configuration where a
detached magnetospheric rope is realized. The blue line is the stellar
surface.}
\end{minipage} 
\end{figure}

All the equilibrium models we obtained are energetically dominated by
the poloidal component of the magnetic field, the energy of the external toroidal magnetic field is,
at most, 25\% of the total magnetic energy of the magnetosphere. This result is similar to what is found when the
twisted field is fully confined within the star. The amount of twist
in the magnetosphere can be regulated essentially by the amount of
poloidal non-linear currents that are imposed to the system.
When the non-linear current terms are weak, the magnetic field lines
are inflated outward by the toroidal magnetic field pressure and the
twist of the field lines extends also to higher latitudes. The result
is a single magnetically connected region, where all field lines have
footpoints attached to the stellar surface. As the currents increase the
effects of the non-linearity get stronger. This not
only increases the twist of the near-surface magnetic field but also
leads to the formation of a disconnected magnetic island, reminiscent
of the so-called plasmoids often found in simulations of the solar
corona. This regime and these topologies are very likely to be
unstable. Only magnetic ropes confined close to the stellar
surface satisfy the  Kruskal-Shafranov condition for stability against
the development of kink.  For all the configurations computed, the internal
linear current is always greater than the external one 
reaching similar values only for configurations where the energy ratio
reaches a maximum.  Apparently, as one tries to rise the external
currents, the system self-regulates inducing a change in the
topology of the distribution of the magnetic field and the associated
external current. As a consequence there is a maximum twist that can
be imposed to the magnetosphere, before reconnection and plasmoid
formation sets in. We found moreover that external currents contribute to the net dipole without affecting too much the strength of the magnetic field at the surface.

Magnetized rapidly rotating neutron stars have been invoked as a
promising engine for both long duration GRBs
\cite{Bucciantini_Quataert+09a,Metzger_Gioannios+11a}, and for short
duration GRBs \cite{Bucciantini_Metzger+12a}. The energy
losses due to the emission of a relativistic magnetically driven  wind
can explain the bulk of known events, but requires in general a high
efficiency to convert the rotational energy of the proto-neutron star
into wind kinetic energy. The presence of a strong magnetic field $\ge
10^{15}$G, is however expected to induce deformations in the neutron
star, that can lead to copious emission of gravitational waves for
rapid rotators \cite{Mastrano_Melatos+11a,Mastrano_Larsky+13a}. The energy loss via gravitational waves will then
compete with the electromagnetic emission of the wind \cite{DallOsso_Stella07a}. It is thus
crucial to evaluate the level of quadrupolar deformation induced by the
magnetic field. We have estimated the magnitude of this quadrupolar
deformation for various cases, both for poloidal and toroidal magnetic
fields, and various current and magnetic field distributions. 

We find that, for a NS with a typical mass of $\sim 1.5 M_\odot$, in
the case of a purely toroidal magnetic field, the quadrupolar
deformation scales as $\|\epsilon_B\|\simeq 5\times 10^{-5}B_{16}^2/m$,  in term of
the maximum value of the magnetic field in units of
$10^{16}$G. Clearly models with a large $m$ where the magnetic field
is concentrated toward the NS surface, can have deformations that are
even an order of magnitude smaller than the case $m=1$ where the field
is more concentrated toward the center. For neutron stars with a purely
poloidal magnetic field, the quadrupolar deformation is found to scale
as $ \|\epsilon_B\|\simeq (1-5)\times 10^{-5}B_{16}^2$, where the upper limit
is for the case of additive currents, with magnetic field concentrated
in the outer stellar layers, and the lower bound for subtractive
currents where the field is concentrated toward the axis, and the bulk
of the star is weakly magnetized. This is the opposite trend with
respect to the one found
for the toroidal case. In our mixed twisted torus configurations, the
deformation is always oblate, and  dominated by the poloidal
field, which is the energetically dominant component. However, such
limitation is probably due to the barotropicity assumption in the
equation of state. If this assumption is relaxed, in principle systems
can be build with a large fraction of magnetic energy into the
toroidal component. This cannot be tested with our algorithm, where
barotropicity is a key assumption. However, if we compare the
deformation derived from purely toroidal field, with the one in the
purely poloidal case, it is evident, that typical twisted torus
configurations, where the toroidal field is confined to small regions
close to the surface (reminiscent of cases with large $m>10$), while
the poloidal component threads the entire star, will likely have a
quadrupolar deformation dominated by the poloidal magnetic field unless the
toroidal component is at least a few times stronger than the poloidal
one. Substantial quadrupolar deformation in this case will likely
require a toroidal field at least an order of magnitude stronger that
the poloidal one.

\section{Conclusions}

We have presented here some recent results in the development of
numerical models for equilibrium configurations of magnetized neutron stars in the fully
non-linear general relativistic regime. This allows us to go beyond the
simplified perturbative regime (where often only the fluid variables
are perturbed while the metric terms are kept fixed), and to derive
the correct absolute scaling that we can then extrapolate to those
regimes more representative of physical situations. The algorithm we have
developed has proved to be robust and efficient, also in handling
extremely deformed objects, giving results that
are in agreement with more sophisticated tecniques. Based on this
approach a publicly available software written in \texttt{FORTRAN90} has been released, together
with visualization and data reduction tools written in \texttt{IDL}, and can be
downloaded at \texttt{www.arcetri.astro.it/science/ahead/XNS/}. Here we have shown how, by choosing different
functional forms for the distribution of currents, it is possible to
realize different magnetic field configurations. We have
investigated how the magnetic field distribution affects the
deformation of the star, in the context of possible gravitational wave
emission from fast rotators. We have shown that, at least within a
barotropic formalism for the equation of state, it is not possible to build configurations of mixed
poloidal-toroidal field, where the toroidal component is energetically
dominant, and that the equation of state itself (in particular the
stratification in the outer part of the neutron star) is
important. Finally we have presented also models where the currents
extend outside the neutron star into a twisted magnetosphere. These
configurations are thought to be more representative of the physical
regime characterizing magnetars. The code we have developed can also
model rotating systems, again both for poloidal and toroidal
configurations (in the latter case it is possible to  model also a
differentially rotating profile) taking into account the fact that in
GR rotating systems have an induced electric field that cannot be
neglected, as it is usually done for non-relativistic MHD. We hope in the future to be able to
evaluate the observable signatures of the magnetic field distributions
that we have computed, either in the form of cyclotron
absorption features, or in the polarized pattern that the emitted
radiation acquires as it travels in the magnetosphere. This will allow
us to check if it is e possible to
constrain the magnetic geometry or maybe even GR effects with
future missions for X-ray polarimetry \cite{Soffitta_Barcons+13a}.

\section*{References}
\bibliography{my}

\end{document}